\documentclass[12pt,draftclsnofoot, peerreview, onecolumn]{IEEEtran}
\usepackage[T1]{fontenc}
\usepackage[latin1]{inputenc}
\usepackage{graphicx}
\usepackage{amssymb}

\providecommand{\tabularnewline}{\\}

\makeatletter

\def\binom#1#2{{#1\choose#2}}

\usepackage[nomarkers]{endfloat}

\makeatother
\begin{document}

\title{Exact and Approximate Expressions for the Probability of Undetected
Error of Varshamov-Tenengol'ts Codes}

\author{Marco Baldi, Franco Chiaraluce,~\IEEEmembership{Member,~IEEE},
 and Torleiv Kl{\o}ve,~\IEEEmembership{Fellow,~IEEE} %
\thanks{The material in this paper was presented in part at the IEEE International
Conference on Communications, Glasgow, Scotland, June 2007.}
\thanks{M. Baldi and F. Chiaraluce are with the Department of
Electronics, Artificial Intelligence and Telecommunications, Università
Politecnica delle Marche, Ancona, Italy (e-mail: \{m.baldi, f.chiaraluce\}@univpm.it).}
\thanks{T. Kl{\o}ve is with the Department of Informatics, University of
Bergen, Bergen, Norway (e-mail: torleiv.klove@ii.uib.no).
His research has been supported by The Norwegian Research Council under grant no. 160236/V30.
}}

\maketitle
\begin{abstract}
Computation of the undetected error probability for error correcting
codes over the Z-channel is an important issue, explored only in part
in previous literature. In this paper we consider the case of Varshamov-Tenengol'ts
codes, by presenting some analytical, numerical, and heuristic methods
for unveiling this additional feature. Possible comparisons with Hamming
codes are also shown and discussed. 
\end{abstract}

\IEEEpeerreviewmaketitle

\section{Introduction}

\label{sec:one}

The Z-channel is a memoryless binary channel. For this channel, a
1 can be changed to a 0 with some probability $p$ (called the channel
error probability), but a 0 is not changed. This channel is a useful
model for a number of applications, like semiconductor memories, some
kinds of optical systems, and other practical environments (examples
can be found in \cite[Chapter 7]{RaoF} and \cite{KE}). In \cite{Zhang},
it was demonstrated that the Z-channel is the only binary-input binary-output
channel that is never dropped in optimal probability loading for parallel
binary channels with a total probability constraint. For a survey
of classical results on codes for the Z-channel, see \cite{Klovebook}.

Several constructions can be adopted for designing codes over the Z-channel 
with given length and error correction capability, and bounds on their size 
can be derived, based on each specific construction \cite{Etzion}. For single
error correcting codes, that are of interest in this paper, further bounds 
can be found in \cite{Shilo}.

We consider a well known class of single error correcting codes for the Z-channel,
that is the class of \emph{Varshamov-Tenengol'ts} (VT) codes \cite{VarshamovT}.
We describe these codes and some of their properties. Let $F_{2}=\{0,1\}$
denote the binary field, and let $Z_{n+1}$ be the additive group
of integers modulo $n+1$. For each $g\in Z_{n+1}$, the VT code $V_{g}$
of length $n$ is the set of vectors ${\mathbf{x}}=(x_{1},x_{2},\ldots,x_{n})\in F_{2}^{n}$
such that 
\begin{equation}
\sum_{m=1}^{n} mx_{m}\equiv g\pmod{n+1}.
\label{eq:eq1}
\end{equation}
 We can observe that the all-zero codeword, noted by ${\mathbf{0}}$,
always belongs to $V_{0}$, while the all-one codeword, noted by ${\mathbf{1}}$,
belongs to $V_{g}$ with $g=\lfloor\frac{n+1}{2}\rfloor$, where $\lfloor x\rfloor$
is the largest integer $m$ such that $m\le x$.

Construction (\ref{eq:eq1}) can be generalized using other abelian
groups of size $n+1$. The corresponding codes are known as Constantin-Rao
(CR) codes. In this paper we only consider VT codes, but most of our
results can easily be generalized to CR codes.

The Hamming weight of ${\mathbf{x}}=(x_{1},x_{2},\ldots,x_{n})\in F_{2}^{n}$
is $w({\mathbf{x}})=\#\{ m\mid x_{m}=1\}$.

We use $\# V_{g}$ to denote the size of $V_{g}$ and $A_{0}^{(g)},A_{1}^{(g)},\ldots,A_{n}^{(g)}$
to denote its weight distribution, that is, $A_{i}^{(g)}$ is the
number of codewords in $V_{g}$ of Hamming weight $i$. Exact formulas
for the size and weight distribution of $V_{g}$ were first determined
by Mazur \cite{Mazur}. They were later generalized to the larger
class of CR codes \cite{HellesethK}. In particular, it is known that
$\# V_{0}\ge\# V_{g}$ for all $g>0$. The codes have all size approximately
$2^{n}/(n+1)$. More precisely, \[
\frac{2^{n}}{n+1}\le\# V_{0}\le\frac{2^{n}}{n+1}\Bigl\{1+\frac{n}{2^{2(n+2)/3}}\Bigr\}.\]
 Further 
\begin{equation}
A_{j}^{(0)}=\frac{1}{n+1}
\sum_{d|(n+1)}(-1)^{j+\lfloor j/d\rfloor}{\frac{n+1}{d}-1 \choose \lfloor j/d\rfloor}\varphi(d)
\label{eq:eq13}
\end{equation}
 where $\varphi(d)$ is the Euler's totient function.

Taking only the main term of (\ref{eq:eq13}) we get the approximation
\begin{equation}
A_{j}^{(0)}\approx\frac{1}{n+1}\binom{n}{j}.
\label{Alapp}
\end{equation}

We let 
${\mathbf{y}}=(y_{1},y_{2},\ldots,y_{n})\le {\mathbf{x}}=(x_{1},x_{2},\ldots,x_{n})$
 denote that $y_{m}\le x_{m}$ for $1\le m\le n$.

When ${\mathbf{x}}$ is sent, then only vectors ${\mathbf{y}}\le{\mathbf{x}}$
can be received, and the probability for this to happen is 
\[
p^{w({\mathbf{x}})-w({\mathbf{y}})}(1-p)^{w({\mathbf{y}})}
=p^{w({\mathbf{e}})}(1-p)^{w({\mathbf{x}})-w({\mathbf{e}})}\]
 where ${\mathbf{e}}={\mathbf{x}}-{\mathbf{y}}$ is the \emph{error
vector}.

A systematic version of VT codes was studied in \cite{AGF}.

Many properties of these codes, either in systematic or non-systematic
form, were explored in the past but, at the best of our knowledge,
no attention has been paid up to now to their error detection properties.

In this paper, we provide a first contribution for filling such gap.
Our analysis is mainly focused on the VT code $V_{0}$. However, we
will also give some results for codes $V_{g}$ with $g\ne0$. Some
comparisons will be also developed with the well known family of Hamming
codes, finding important performance similarities, both when these
codes are applied over the Z-channel and even over the symmetric channel.

The paper is organized as follows. In Section \ref{sec:two} we introduce
the probability of undetected error, $P_{\mathrm{u}e}$. In Section
\ref{sec:three} we give an exact formula for $P_{\mathrm{u}e}$,
that can be explicitly computed for small lengths $n$ (up to approximately
25). Next, in Section \ref{sec:four} we study good lower bounds that
can be explicitly computed up to almost twice this length (depending
on how tight we require the bounds to be). In Section \ref{sec:fourbis}
we look at the class of Hamming codes, for the sake of comparison,
and their application is considered for both the symmetric channel
and the asymmetric one; a first performance comparison with VT codes
is done, for small lengths. In Section \ref{sec:five} we use some
heuristic arguments to give a very good approximation that can easily
be computed even for large lengths. In Section \ref{sec:six} we use
Monte Carlo methods to obtain other good approximations for long code
lengths; this permits us also to make other comparisons with Hamming
codes of the same length. Finally, in Section \ref{sec:seven}, some
remarks on future research conclude the paper.

\section{The probability of undetected error}

\label{sec:two} For a description of properties of the probability
of undetected error, see \cite{K}. In general, an undetected error
occurs when, in presence of one or more errors, the received sequence
coincides with a codeword different from the transmitted one. In this
case the decoder accepts the received sequence, and information reconstruction
is certainly wrong. By the VT code construction, single errors are
always detected, so that undetected errors can appear only when the
number of errors is greater than one.

We note that, if ${\mathbf{x}}\in V_{g}$ is sent and ${\mathbf{y}}$
is received, then ${\mathbf{y}}\in V_{g}$ if and only if ${\mathbf{e}}={\mathbf{x}}-{\mathbf{y}}\in V_{0}$.
This can be proved by observing that: 
\[\sum_{m=1}^{n} me_{m}=\sum_{m=1}^{n} mx_{m}-\sum_{m=1}^{n} my_{m}\equiv g-g=0\pmod{n+1}.\]
 Hence, the undetectable errors are exactly the non-zero vectors in $V_{0}$.
For $j\ge 0$, let \[
{\cal {E}}_{j}({\mathbf{x}})=\{{\mathbf{e}}\in V_{0}\mid w({\mathbf{e}})=j,\ {\mathbf{e}}\le{\mathbf{x}}\}.\]
For $j>0$, this is the set of undetectable errors of weight $j$ when ${\mathbf{x}}$
is transmitted. We note that ${\cal {E}}_{0}({\mathbf{x}})=\{{\mathbf{0}}\}$ 
(and ${\mathbf{0}}$ is not an error vector).
Let $\varepsilon_{j}({\mathbf{x}})$ be the size
of ${\cal {E}}_{j}({\mathbf{x}})$. Note that since $V_{0}$ does
not contain any vector of weight one, we have $\varepsilon_{1}({\mathbf{x}})=0$
for all ${\mathbf{x}}$. We also have $\varepsilon_{0}({\mathbf{x}})=1$
for all ${\mathbf{x}}$.

The (average) probability of undetected error is given by \begin{equation}
P_{\mathrm{u}e}(V_{g},p)=\frac{1}{\# V_{g}}\sum_{{\mathbf{x}}\in V_{g}}\sum_{j=2}^{w({\mathbf{x}})}\varepsilon_{j}({\mathbf{x}})p^{j}(1-p)^{w({\mathbf{x}})-j}.\label{eq:eq2}\end{equation}
 In deriving (\ref{eq:eq2}), the codewords are assumed equally
probable. If we define \[
A_{i,j}^{(g)}=\sum_{{{\mathbf{x}}\in V_{g}\atop w({\mathbf{x}})=i}}\varepsilon_{j}({\mathbf{x}}),\]
 (\ref{eq:eq2}) can be rewritten \begin{equation}
P_{\mathrm{u}e}(V_{g},p)=\frac{1}{\# V_{g}}\sum_{i=2}^{n}\sum_{j=2}^{i}A_{i,j}^{(g)}p^{j}(1-p)^{i-j}.\label{eq:eq3}\end{equation}

\section{Exact evaluation of the undetected error probability}

\label{sec:three}

Conceptually, the simplest way to compute the undetected error probability
consists in direct calculation of (\ref{eq:eq2}) by first determining
the sets ${\cal {E}}_{j}({\mathbf{x}})$. Since both $V_{g}$ and
$V_{0}$ have size on the order of $2^{n}/(n+1)$, the complexity
is on the order of $2^{2n}/(n+1)^{2}$.

\subsection{$P_{\mathrm{u}e}(V_{0},p)$}

We observe that if ${\mathbf{x}}=(x_{1},x_{2},\ldots,x_{n})\in V_{0}$,
then the reversed vector \[
{\mathbf{x}}^{\rho}=(x_{n},x_{n-1},\ldots,x_{1})\in V_{0},\]
 too, since \begin{eqnarray*}
\sum_{m=1}^{n} mx_{n+1-m} & = & \sum_{m=1}^{n}(n+1-m)x_{m}\\
 & = & w({\mathbf{x}})(n+1)-\sum_{m=1}^{n} mx_{m}\\
 & \equiv & 0-0=0\pmod{n+1}.\end{eqnarray*}
 This simplifies the calculations and reduces the complexity by some
factor, but the order of magnitude of the complexity is still the
same. We will elaborate on this symmetry in the next section.

Another observation is that if $w({\mathbf{x}})=i$ and ${\mathbf{e}}\in{\cal {E}}_{j}({\mathbf{x}})$,
with $j\le i$, then ${\mathbf{y}}\in{\cal {E}}_{i-j}({\mathbf{x}})$.
Hence, 
\begin{equation}
A_{i,j}^{(0)}=A_{i,i-j}^{(0)}\mbox{ for }0\le j\le i\le n.
\label{sym}
\end{equation}
This again halves the complexity for $g=0$.
For completeness, we also observe that 
\begin{equation}
A_{i,i}^{(0)}=A_{i}^{(0)}\mbox{ for }0\le i\le n,
\label{sym2}
\end{equation}
\begin{equation}
A_{i,1}^{(0)}=0\mbox{ for }1\le i\le n.
\label{sym3}
\end{equation}

For even $n$, further symmetry properties can be found.

For any vector $\mathbf{x}=(x_1,x_2,\ldots ,x_n)$, the \emph{complementary vector} $\bar{\mathbf{x}}$
is defined by 
\[\bar{\mathbf{x}}=(1-x_1,1-x_2,\ldots ,1-x_n),\]
that is, $\bar{x}_m=1$ if $x_m=0$ and $\bar{x}_m=0$ if $x_m=1$.
Clearly, 
\[w(\bar{\mathbf{x}})=n-w(\mathbf{x})\]
and
\[\sum_{m=1}^n m \bar{x}_m = \frac{n^2(n+1)}{2}-\sum_{m=1}^n m x_m.\]

This implies that if $n$ is even and ${\mathbf{x}}\in V_0$,
then $\bar{\mathbf{x}}\in V_0$. 
Next, we observe that if ${\mathbf{y}}\le {\mathbf{x}}$, then $\bar{\mathbf{x}}\le \bar{\mathbf{y}}$.
In particular, this implies the relation
\begin{equation}
\label{rel1}
A_{i,j}^{(0)}=A_{n-j,n-i}^{(0)}\mbox{ for }0\le j\le i\le n.
\end{equation}

We note that this relation is not valid for odd $n$.

Relations (\ref{sym}) and (\ref{rel1}) can be combined. For example, (\ref{rel1}) implies that
$A_{i,i-j}^{(0)}=A_{n-i+j,n-i}^{(0)}$. Next, (\ref{sym}) implies that $A_{n-i+j,n-i}^{(0)}=A_{n-i+j,j}^{(0)}$,
etc. Repeated use of (\ref{sym}) and (\ref{rel1}) gives the following result:

if $n$ is even and $0\le j\le i \le n$, then
\begin{eqnarray}
A_{i,j}^{(0)}&=& A_{i,i-j}^{(0)}=A_{n-i+j,n-i}^{(0)}=A_{n-i+j,j}^{(0)}\nonumber \\
       &=& A_{n-j,i-j}^{(0)}=A_{n-j,n-i}^{(0)}.
\label{rel2}
\end{eqnarray}

Putting $i=j$ in (\ref{rel2}) and combining with 
 (\ref{sym2}) we also get
\begin{equation}
A_{n,j}^{(0)}=A_{j}^{(0)}=A_{n-j}^{(0)}.
\label{sym4}
\end{equation}

Using these relations, for even $n$, the complexity of the exact calculus for $g=0$
is further reduced. For odd $n$, the same relationships are
not valid.
Moreover, it should be noted that, for odd $n$ we have $A_{n}^{(0)}=A_{n,j}^{(0)}=0$.

We have developed a numerical program, in the C++ language, that constructs
all the sets ${\cal {E}}_{j}({\mathbf{x}})$ for ${\mathbf{x}}\in V_{0}$
(exploiting the symmetry properties discussed above) and, based on
this, computes the numbers $A_{i,j}^{(0)}$. The values of $P_{{\rm ue}}(V_{0},p)$
as a function of $p$ computed this way are exact.

\begin{figure}[ht]
 \centering \includegraphics[scale=0.65]{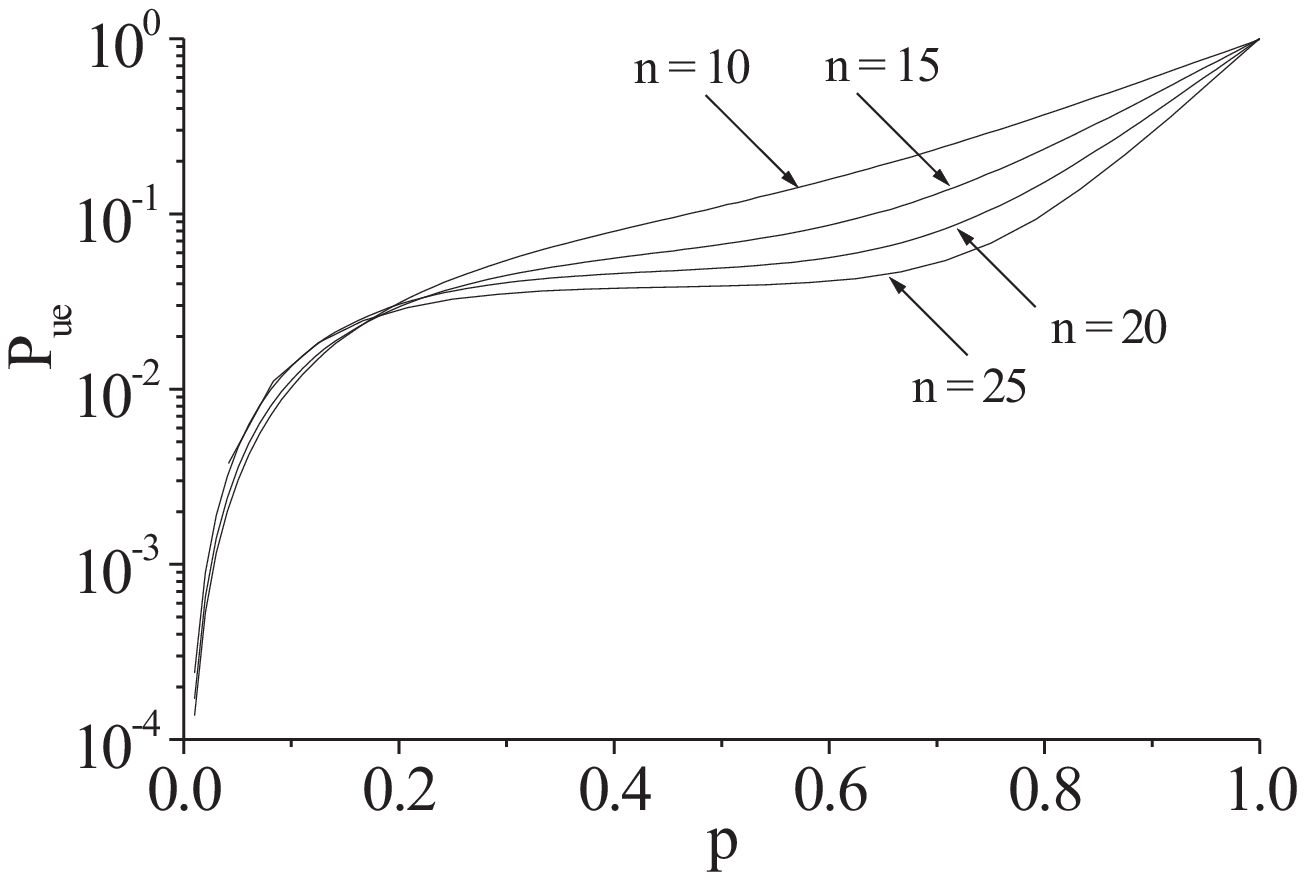}

\caption{$P_{{\rm ue}}(V_{0},p)$ vs. $p$ for some values of $n$.}

\label{fig:Fig1} 
\end{figure}

Examples of the results obtained are shown in Fig.~\ref{fig:Fig1},
for $n=10,15,20,25$. For small values of $p$, up to about 0.2, small
values of $n$ give lower probability of undetected error, whilst
for larger values of $p$ the behavior of codes with larger $n$ is
better.

For $p=1$ all codewords are changed to ${\mathbf{0}}$, and this
implies: \[
P_{{\rm ue}}(V_{0},1)=\frac{\# V_{0}-1}{\# V_{0}},\]
 that is slightly different from 1 because of the presence of the
all-zero codeword (that is always received correctly).

\subsection{$P_{\mathrm{u}e}(V_{g},p)$ for $g\ne0$}

\label{subsec:PueVg} For $g\ne0$, $V_{g}$ does not include the
all-zero vector. As a consequence, $P_{\mathrm{u}e}(V_{g},1)=0$ and
the curve of $P_{{\rm ue}}$ has at least one maximum for $p$ between
0 and 1. As for $V_{0}$, we have developed a computer program that
permits us to calculate exactly the undetected error probability of
these codes, as a function of $p$, for not too high values of $n$
(in such a way as to have acceptable processing times).

\begin{figure}[ht]
 \centering \includegraphics[scale=0.65]{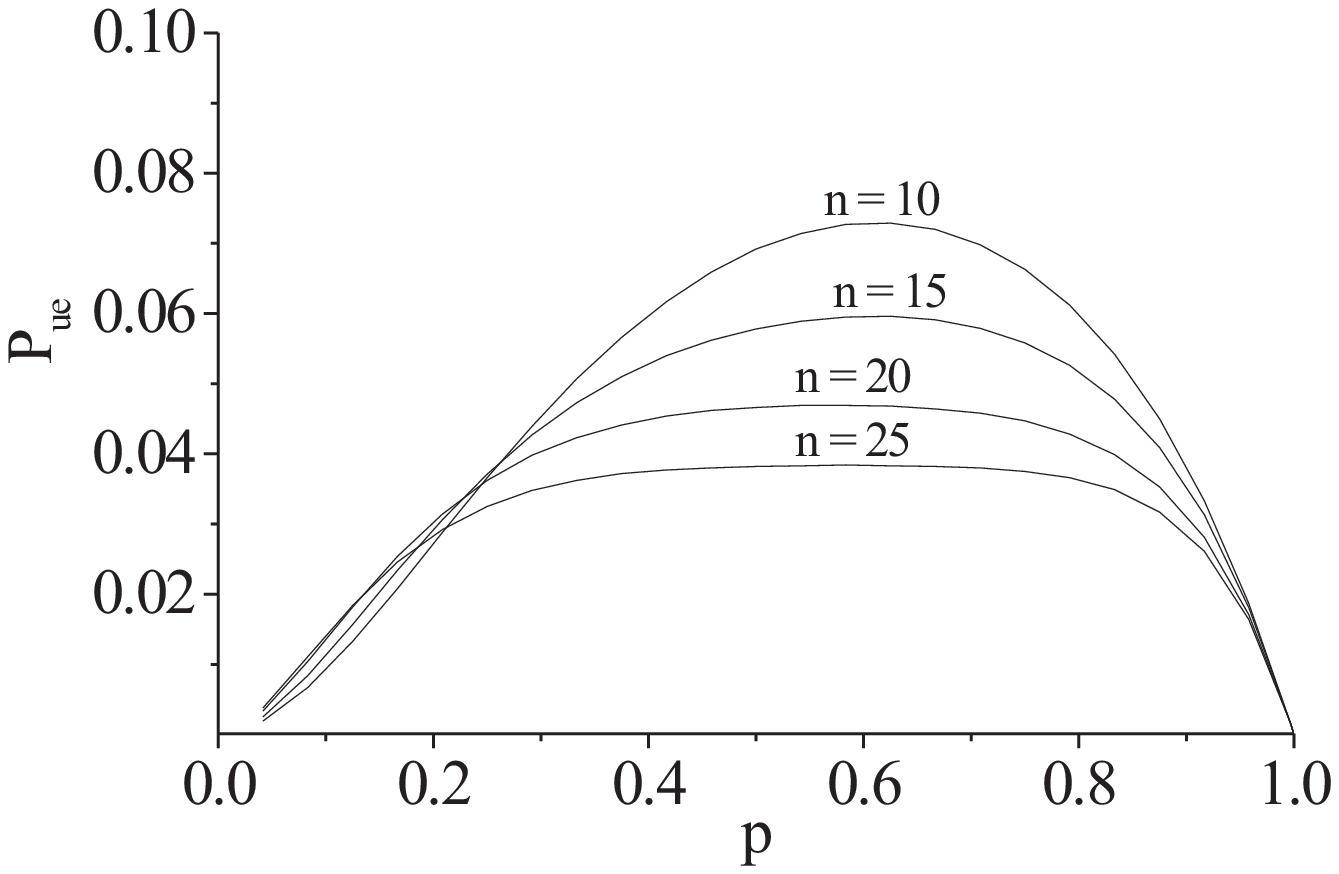}

\caption{$P_{{\rm ue}}(V_{1},p)$ vs. $p$ for some values of $n$.}

\label{fig:Fig2} 
\end{figure}

In Fig. \ref{fig:Fig2}, curves of $P_{\mathrm{u}e}(V_{1},p)$ are
plotted, for some values of $n$. For better readability, we have
used a linear scale instead of the logarithmic one used in Fig. \ref{fig:Fig1}.

These curves have been obtained for $g=1$, but they remain practically
the same for codes $V_{g}$, with $g>1$.

\begin{figure}[ht]
 \centering \includegraphics[scale=0.75]{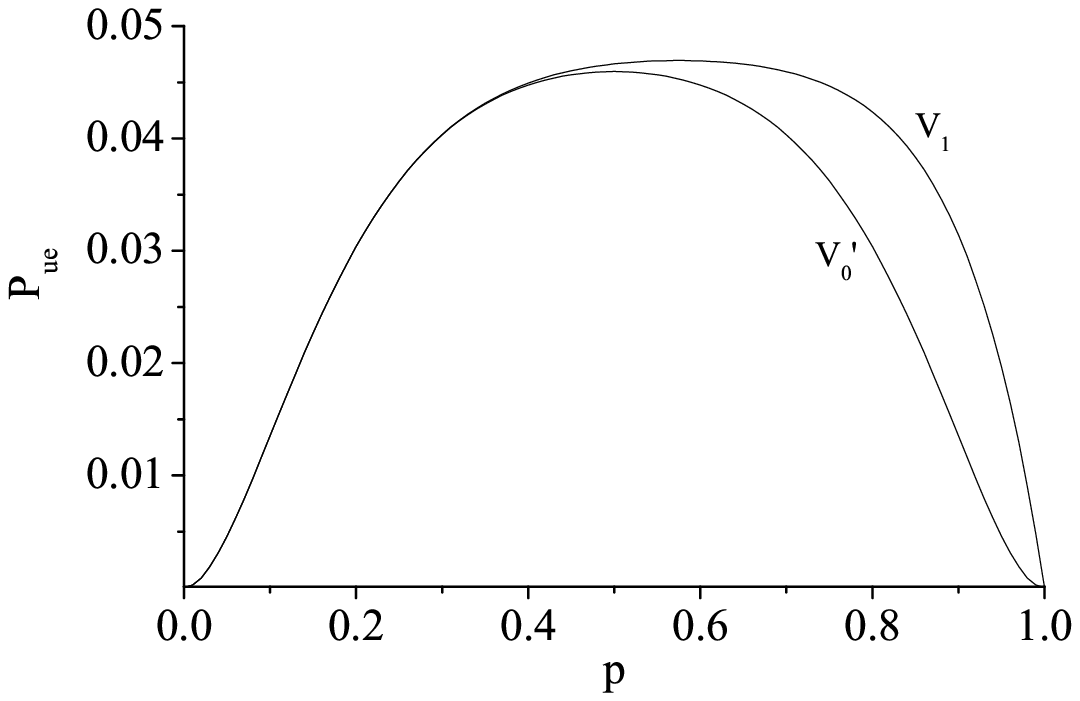}

\caption{Comparison of $P_{{\rm ue}}(V_{0}',p)$ and $P_{{\rm ue}}(V_{1},p)$
for $n=20$.}

\label{fig:Fig10} 
\end{figure}

A main reason why $P_{\mathrm{u}e}(V_{0},p)$ and $P_{\mathrm{u}e}(V_{1},p)$
behave differently for large $p$ is that $V_{0}$ contains the all-zero
vector. If we remove the all-zero codeword, that is consider 
\[V_{0}'=V_{0}\setminus \{{\mathbf{0}} \}\]
instead, we get: \[
P_{{\rm ue}}(V_{0}',p)=P_{{\rm ue}}(V_{0},p)-\frac{\sum_{j=1}^{n}A_{j}^{(0)}p^{j}}{\# V_{0}}\]
 and code $V_{0}'$ has no undetectable errors for $p=1$. 
It turns out that $P_{\mathrm{u}e}(V_{0}',p)$ and $P_{\mathrm{u}e}(V_{1},p)$
are almost the same for $p<0.5$ but they differ somewhat in the region
$[0.5,1]$. We illustrate this behavior for $n=20$, in Fig. \ref{fig:Fig10}.

\section{Lower bounds on $P_{{\rm ue}}$}

\label{sec:four} Clearly, if we omit or reduce some of the terms
in (\ref{eq:eq3}), we get a lower bound on $P_{\mathrm{u}e}(V_{g},p)$.
For example, for a fixed integer $m\geq2$, then \[
P_{\mathrm{u}e}(V_{g},p)\ge\frac{1}{\# V_{g}}\sum_{i=2}^{n}\sum_{j=2}^{\min(m,i)}A_{i,j}^{(g)}p^{j}(1-p)^{i-j}.\]

The number of errors of weight $j$ is upper bounded by $A_{j}^{(0)}\approx\frac{1}{n+1}\binom{n}{j}$.
Hence, the complexity of calculating the coefficients $A_{i,j}^{(g)}$
for $j\le m$ is on the order of \[
\frac{2^{n}\binom{n}{m}}{(n+1)^{2}}.\]
 For small values of $m$, this is of course much lower than the computations
needed to determine all the $A_{i,j}^{(g)}$ which we estimated to
be on the order of $2^{2n}/(n+1)^{2}$.

Next, we describe in detail how to calculate $A_{i,j}^{(0)}$ for
$j=2$, $j=3$ and $j=4$. First, we remind the reader that the support $\chi({\mathbf{e}})$ of a vector ${\mathbf{e}}$ is the set of positions where the vector has ones, that is
\[\chi({\mathbf{e}})=\{r \mid e_r=1\}.\]

\begin{itemize}
\item Calculus of $A_{i,2}^{(0)}$ 
\end{itemize}
If ${\mathbf{e}}\in V_{0}'$ has weight 2, and \[
\chi({\mathbf{e}})=\{ r,s\},\]
where $1\le r<s\le n$, then, by the definition of the code, we must have
\[
r+s=n+1.\]
 Hence $s=n+1-r\ge r+1$ and so $r\le n/2$. Therefore, ${\mathbf{e}}\in{\cal {E}}_{2}({\mathbf{x}})$
if and only if $x_{r}=x_{n+1-r}=1$. Hence \[
\varepsilon_{2}({\mathbf{x}})=\sum_{r=1}^{\lfloor n/2\rfloor}x_{r}x_{n+1-r},\]
 and \[
A_{i,2}^{(0)}=\sum_{{{\mathbf{x}}\in V_{0}'\atop w({\mathbf{x}})=i}}\sum_{r=1}^{\lfloor n/2\rfloor}x_{r}x_{n+1-r}.\]

\begin{itemize}
\item Calculus of $A_{i,3}^{(0)}$ 
\end{itemize}
If ${\mathbf{e}}\in V_{0}'$ has weight 3, and \[
\chi({\mathbf{e}})=\{ r,s,t\}\]
 where $1\le r<s<t\le n$, then we must have \[
r+s+t=n+1\mbox{ or }r+s+t=2(n+1).\]
 We observe that if ${\mathbf{e}}\le{\mathbf{x}}$ then, clearly,
${\mathbf{e}}^{\rho}\le{\mathbf{x}}^{\rho}$ (the reversed vectors).
Further, \[
\chi({\mathbf{e}}^{\rho})=\{ n+1-t,n+1-s,n+1-r\}\]
 and \[
(n+1-t)+(n+1-s)+(n+1-r)=3(n+1)-(r+s+t).\]
 Hence, for each error with support that sums to $n+1$, there is
another (reversed) error with support that sums to $2(n+1)$. Therefore,
it is sufficient to consider the first kind, this way deriving a contribution
that is exactly half of $A_{i,3}^{(0)}$. If $r+s+t=n+1$, then we
have \[
n+1=r+s+t\ge r+(r+1)+(r+2)\]
 and so $r\le(n-2)/3$. Further, \[
n+1-r=s+t\ge s+(s+1)\]
 and so $s\le(n-r)/2$. Hence, similarly to what we did for $j=2$,
we get

\[
A_{i,3}^{(0)}=2\sum_{{{\mathbf{x}}\in V_{0}'\atop w({\mathbf{x}})=i}}\sum_{r=1}^{\lfloor(n-2)/3\rfloor}\sum_{s=r+1}^{\lfloor(n-r)/2\rfloor}x_{r}x_{s}x_{n+1-r-s}.\]

\begin{itemize}
\item Calculus of $A_{i,4}^{(0)}$ 
\end{itemize}
If ${\mathbf{e}}\in V_{0}'$ has weight 4, and \[
\chi({\mathbf{e}})=\{ r,s,t,u\}\]
 where $1\le r<s<t<u\le n$, then one of the following conditions
should be satisfied: \def\theenumi{\roman{enumi}}

\begin{enumerate}
\item $r+s+t+u=n+1$, 
\item $r+s+t+u=2(n+1)$, 
\item $r+s+t+u=3(n+1)$. 
\end{enumerate}
However, we observe that when vector $\mathbf{e}$ satisfies condition
i) then the reversed vector ${\mathbf{e}}^{\rho}$ satisfies condition
iii) and vice versa. In fact: \begin{eqnarray*}
(n+1-u)+(n+1-t)+(n+1-s)+(n+1-r)=\\
=4(n+1)-(r+s+t+u)=3(n+1).\end{eqnarray*}
 Hence, for each error with support that sums to $n+1$, there is
another (reversed) error with support that sums to $3(n+1)$. Therefore,
it is sufficient to consider condition i) and then double the size
so found for taking into account also condition iii). If $r+s+t+u=n+1$,
we have \[
n+1=r+s+t+u\ge r+(r+1)+(r+2)+(r+3)=4r+6\]
 and so $r\le(n-5)/4$. On the other hand, \[
n+1-r=s+t+u\ge s+(s+1)+(s+2)=3s+3\]
 and so $s\le(n-2-r)/3$. Finally \[
n+1-r-s=t+u\ge t+(t+1)=2t+1\]
 and so $t\le(n-r-s)/2$.

Similarly, we can consider condition ii). It implies: \[
2n+2=r+s+t+u\ge4r+6\]
 and so $r\le(2n-4)/4$. Further \[
2n+2-r=s+t+u\ge3s+3\]
 and so $s\le(2n-1-r)/3$. Finally \[
2n+2-r-s=t+u\ge2t+1\]
 and so $t\le(2n+1-r-s)/2$.

On the basis of such analysis, the expression of $A_{i,4}^{(0)}$
can be written as follows: \begin{eqnarray*}
A_{i,4}^{(0)} & = & 2\sum_{{{\mathbf{x}}\in V_{0}'\atop w({\mathbf{x}})=i}}\sum_{r=1}^{\lfloor(n-5)/4\rfloor}\sum_{s=r+1}^{\lfloor(n-2-r)/3\rfloor}\\
 &  & \quad\sum_{t=s+1}^{\lfloor(n-r-s)/2\rfloor}x_{r}x_{s}x_{t}x_{n+1-r-s-t}\\
 &  & +\sum_{{{\mathbf{x}}\in V_{0}'\atop w({\mathbf{x}})=i}}\sum_{r=1}^{\lfloor(2n-4)/4\rfloor}\sum_{s=r+1}^{\lfloor(2n-1-r)/3\rfloor}\\
 &  & \quad\sum_{t=\max(s+1,n+2-r-s)}^{\lfloor(2n+1-r-s)/2\rfloor}x_{r}x_{s}x_{t}x_{2n+2-r-s-t}.\end{eqnarray*}
 It should be noted that, in the inner sum of the second contribution
(the one due to condition ii)), we have explicitly taken into account
that $t$ cannot be smaller than $n+2-r-s$; this is because the following
obvious condition must be satisfied \[
u=2(n+1)-(r+s+t)\le n\]
 that implies \[
n+2-r-s\le t.\]

Additionally, the sums appearing in the expressions of $A_{i,j}^{(0)}$
are null when the upper extreme is smaller than the lower extreme.
So, the first contribution in $A_{i,4}^{(0)}$ is not present for
$n\leq8$, and also the second contribution disappears (as obvious)
for $n\leq3$.

Though the procedure adopted to derive $A_{i,2}^{(0)}$, $A_{i,3}^{(0)}$
and $A_{i,4}^{(0)}$ is quite clear and, in principle, can be extended
to the other values of $j$, it is easy to see that, formally, the
analysis becomes more and more tedious for increasing $j$. Similarly,
explicit formulas can be given for $A_{i,j}^{(g)}$ for $g\ne0$,
but they are usually somewhat more complicated. The formula for $j=2$
generalizes immediately to \[
A_{i,2}^{(g)}=\sum_{{{\mathbf{x}}\in V_{g}\atop w({\mathbf{x}})=i}}\sum_{r=1}^{\lfloor n/2\rfloor}x_{r}x_{n+1-r}.\]
 However, for $j=3$ we used above the symmetry that only appears
in $V_{0}$, and so the formula for $A_{i,3}^{(g)}$ will contain
two sums in the expression. Similarly for $j\ge4$.

\begin{figure}[ht]
 \centering \includegraphics[scale=0.75]{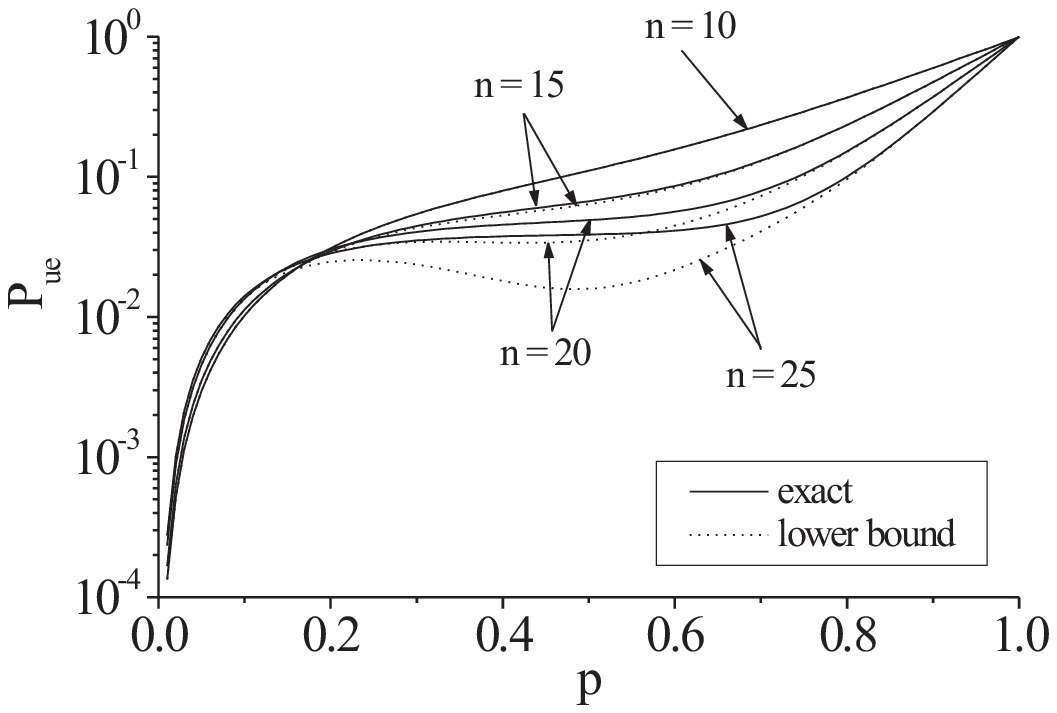}

\caption{The lower bounds and the exact results for $P_{\mathrm{u}e}(V_{0},p)$
for lengths $n=10,15,20,25$.}

\label{fig:Fig4} 
\end{figure}

For $V_{0}$ we have computed some lower bounds for $n=10,15,20$
and $25$ to see how good the bounds are, in comparison with the exact
values. In the lower bound we have used $A_{i,2}$, $A_{i,3}$, and
$A_{i,4}$ for $4\le i\le n$ computed by the formulas above, $A_{i,i-2}$,
$A_{i,i-3}$, $A_{i,i-4}$ which have the same values by (\ref{sym}),
and finally $A_{i,i}$ obtained from (\ref{sym2}). The remaining
terms have been set to zero. The lower bounds and the exact values
are compared in Fig. \ref{fig:Fig4}.

From the figure we see that the lower bound is an excellent approximation
of the true behavior for $n=10$ (the exact curve and the bound are
superposed), that the approximation is very good for $n=15$, but
that the difference between the exact curve and the estimated one
becomes more and more evident for increasing $n$. Qualitatively,
such a trend seems quite obvious and expected. In particular, the
lower bound for $n=25$ exhibits an oscillation, in the central region,
which is due to the terms neglected, whose effect is particularly
important in the neighborhood of $p=0.5$. On the other hand, it is
easy to verify that the approximation is very good, independently
of $n$, for small values of the channel error probability $p$. Even
the simple bound using only $A_{i,2}^{(0)}$ gives a good approximation
for small $p$.

\section{Comparison with Hamming codes and relationship with the symmetric
channel}

\label{sec:fourbis} Hamming codes are another well known class of
single error correcting codes, widely used both in symmetric and asymmetric
channels. In particular, they are known to be optimal error detecting
codes for the the binary symmetric channel (BSC) \cite{KU}.

The length of a binary Hamming code $H$ is $n=2^{r}-1$, where $r$
is the number of parity check bits, while the number of codewords
(i.e., the size of the code) is $M=2^{k}$, with $k=2^{r}-1-r$. For
a description of Hamming codes and their properties see, for example,
\cite{Linbook}.

The dual codes of Hamming codes are maximal length (or simplex) codes,
which means that the generator matrix of a Hamming code can be used
as the parity check matrix of a maximal length code, and vice versa.

The weight distribution of these codes is known: 
\[
A_{i}^{H}=\frac{\binom{n}{i}+n(-1)^{\lceil i/2\rceil }\binom{(n-1)/2}{\lfloor i/2\rfloor}}{n+1}.
\]
Here $\lceil x\rceil$ denotes the smallest integer $m$
such that $m\ge x$.

When the code is applied over the BSC, this permits to find an explicit
expression for the probability of undetected error \cite[p. 44]{K},
namely: 
\begin{equation}
P_{{\rm ue}}^{{\rm BSC}}(H,p)=\frac{1}{n+1}\left[1+n(1-2p)^{(n+1)/2}\right]-(1-p)^{n}.
\label{PueHBSC}
\end{equation}
 In this expression, $p$ represents either the probability that a
1 is changed to a 0 or a 0 is changed to a 1.

However, as for VT codes, an explicit expression for $P_{{\rm ue}}(H,p)$
is not available for the case of the Z-channel. Similarly to what
was done in Section \ref{sec:three}, we have developed a numerical
program, in C++ language, that permits to evaluate, exhaustively,
all transitions yielding undetected errors. The procedure is conceptually
similar to that described in Section \ref{sec:two}, for VT codes,
and an expression like (\ref{eq:eq3}) still holds, as an undetected
error occurs if and only if the error vector belongs to $H$.

\begin{figure}[ht]
 \centering \includegraphics[scale=0.75]{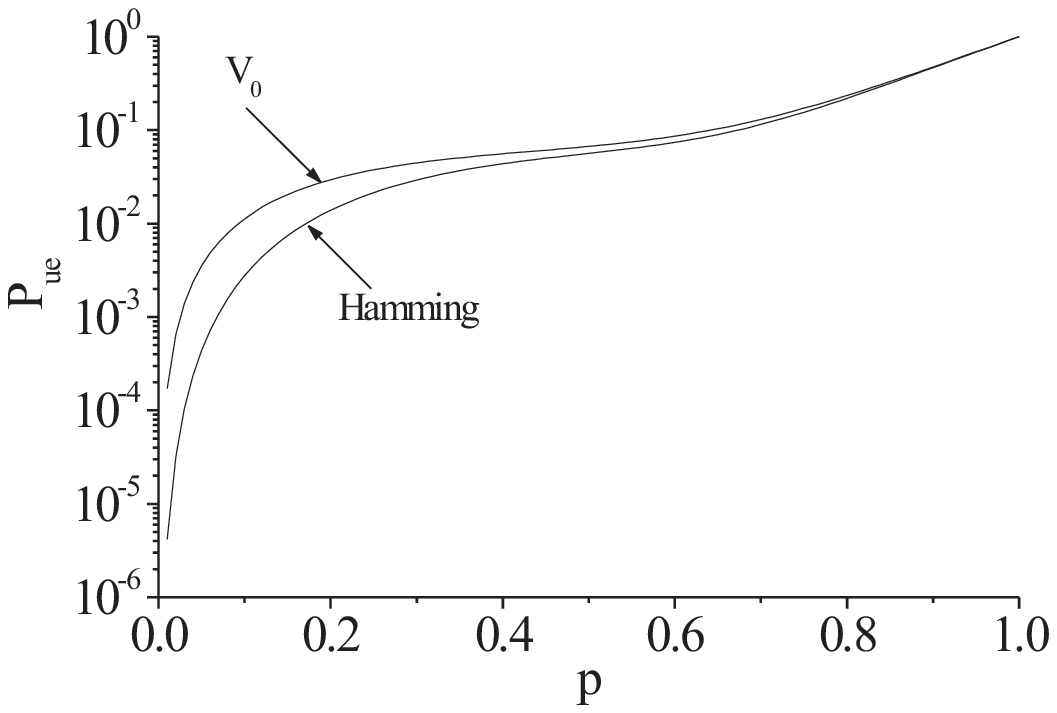}

\caption{Comparison of $P_{{\rm ue}}(V_{0},p)$ and $P_{{\rm ue}}(H,p)$ for
$n=15$.}

\label{fig:HvsV015} 
\end{figure}

The curve of $P_{{\rm ue}}(H,p)$ can be compared, for a fixed $n$,
with that of $P_{{\rm ue}}(V_{0},p)$. An example is shown in Fig.
\ref{fig:HvsV015} for $n=15$; both codes have the same number of
codewords, i.e., $\# V_{0}=M=2048$. The two curves are rather similar,
but the performance of the Hamming code is slightly better. In Section
\ref{sec:six} we will do a comparison for a larger $n$. 
There we show that for $n=127$ both curves are dominated by a nearly flat region
in the neighborhood of $p=0.5$. The extent of the nearly flat region becomes
wider and wider for increasing $n$. The rationale for the existence
of the nearly flat region in the curve of $P_{{\rm ue}}$ is given in the
next section.

\section{Heuristic approximations}

\label{sec:five} The lower bound discussed in the previous section
neglects all the events caused by $j$ errors where $5\leq j\leq i-5$.
As a consequence, the approximation is good for small $p$ (and, symmetrically,
for large $p$) but it becomes less and less reliable in the central
region of $p$ values.

Another approach is to find some good approximation of $A_{i,j}^{(0)}$
by some heuristic argument. By (\ref{sym})-(\ref{sym3}), we only have to
consider $j$ in the range $2\le j\le i/2$.

First, we observe that a vector ${\mathbf{e}}$ of weight $j$ is
contained in $\binom{n-j}{i-j}$ vectors ${\mathbf{x}}$ of weight
$i$. Each such vector ${\mathbf{x}}$ is contained in some code $V_{g}$.
Since there are $A_{j}^{(0)}$ vectors ${\mathbf{e}}\in V_{0}'$ of
weight $j$, we get 
\begin{equation}
\binom{n-j}{i-j}A_{j}^{(0)}=\sum_{g=0}^{n}A_{i,j}^{(g)}.
\label{all}
\end{equation}

Now (and this is the heuristic argument), we assume that the ratio
between the number of undetectable errors of weight $j$ in $V_{g}$
and the overall number of errors of weight $j$ (given by (\ref{all})),
starting from codewords ${\mathbf{x}}$ of weight $i$, is approximately
equal to the ratio between the number of codewords of weight $i$
in $V_{g}$ and the total number of codewords of weight $i$. In particular,
for $V_{0}$, this means to assume: \[
\frac{A_{i,j}^{(0)}}{\binom{n-j}{i-j}A_{j}^{(0)}}\approx\frac{A_{i}^{(0)}}{\binom{n}{i}}.\]
 Hence, under our assumption, we get 
\begin{equation}
A_{i,j}^{(0)}\approx\frac{\binom{n-j}{i-j}}{\binom{n}{i}}A_{j}^{(0)}A_{i}^{(0)}.
\label{app}
\end{equation}
 This approximation can be computed using (\ref{eq:eq13}).
We observe that for $i=n$, we have equality in (\ref{app}), that is, 
\[A_{n,j}^{(0)}=A_{j}^{(0)}A_{n}^{(0)}.\]

Even more simply, as an alternative to using (\ref{eq:eq13}),
one can combine (\ref{app}) with the approximations for $A_{i}^{(0)}$
and $A_{j}^{(0)}$ given by (\ref{Alapp}) and get 
\begin{equation}
A_{i,j}^{(0)}\approx\frac{1}{(n+1)^{2}}\binom{n-j}{i-j}\binom{n}{j}=\frac{1}{(n+1)^{2}}\binom{n}{i}\binom{i}{j}
\label{app2}
\end{equation}
 for $2\le j\le i-2$ while, using (\ref{sym2}), we get \begin{equation}
A_{i,i}^{(0)}\approx\frac{1}{(n+1)}\binom{n}{i}.\label{app2a}\end{equation}
 Finally, in the case of even $n$, using (\ref{sym4}), we get
\begin{equation}
A_{n,j}^{(0)}\approx\frac{1}{(n+1)}\binom{n}{j}.\label{app2b}\end{equation}

\begin{figure}[ht]
 \centering \includegraphics[scale=0.75]{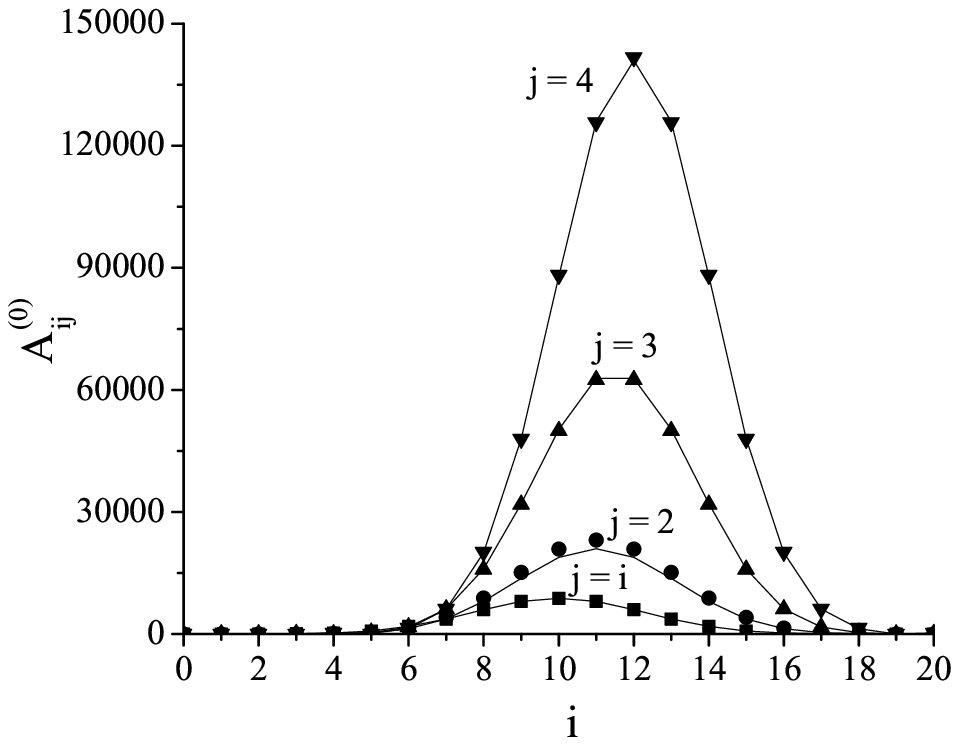}

\caption{Comparison between the exact values of $A_{i,j}^{(0)}$ (dots) and
those derived from the heuristic approximation (continuous lines)
as a function of $i$ for $n=20$ and some values of $j$.}

\label{fig:FigHeur} 
\end{figure}

The heuristic argument is justified by a number of simulation evidences.
Just as an example, in Fig. \ref{fig:FigHeur} we show the comparison
between the exact values of $A_{i,j}^{(0)}$ and those derived from
the heuristic approximation, as a function of $i$, for $n=20$ and
some values of $j$, namely $j=i$ (i.e., using (\ref{app2a}) for
the heuristic approximation), and $j=2,3,4$ ((i.e., using (\ref{app2})
and (\ref{app2b}) for the heuristic approximation). The heuristic
values have been interpolated by continuous lines for the sake of
readability. The figure shows that the agreement between the approximated
values and the exact ones is very good. Though referred to a particular
case, this conclusion is quite general, and we have verified it also
for the other values of $j$ and for different $n$ (for example,
$n=25$). From a theoretical point of view, the heuristic argument
can be seen as an instance of the {}``random coding'' approach,
that has been also used recently, over the Z-channel, to extend the
concept of Maximum Likelihood decoding \cite{Barbero2006}. As the
practical significance of random coding increases with the size of
the code, we can foresee that the goodness of the heuristic argument
is confirmed for larger values of $n$.

In practice, the best approach for moderate $n$ is to determine some
$A_{i,j}^{(0)}$ explicitly, as outlined above, e.g. for $j\le4$,
combined with (\ref{sym}) and (\ref{sym2}), and to use one
of the approximations in (\ref{app}) and (\ref{app2}) for
the remaining $A_{i,j}^{(0)}$. For $n=25$ we have done this, with
exact values for $A_{i,j}^{(0)}$ and $A_{i,i-j}^{(0)}$ for $0\le j\le4$,
and with the approximation in (\ref{app}) for the remaining $A_{i,j}^{(0)}$.
For this case, the exact values and the approximations are very close,
and if we draw both in a graph it is not possible to distinguish between
them. The maximal percentage difference between the curves is less
than 0.065\%; the maximum occurs in the neighborhood of $p=0.5$.

For our heuristic approximation given in terms of the binomial coefficients,
we can find a closed formula. The analytical details are given in
Appendix I; using the approximations in (\ref{app2}) and (\ref{app2a}) we get the following expression:

\begin{eqnarray}
\lefteqn{\left(n+1\right)^{2}\# V_{0} P_{\mathrm{ue}}^{\mathrm{h}}\left(V_{0},p\right)} \nonumber\\
&=&  2^{n}-(2-p)^{n}-np(2-p)^{n-1} \nonumber\\
&& + 2np(1+p)^{n-1}-2np-n(n-1)p^{2}. 
\label{eq12}
\end{eqnarray}

It is easy to see that, for $n$ sufficiently large and except for
$p$ close to zero or one, at the right side, the first term is much
larger than the others. So, taking into account that $(n+1)^{2}\# V_{0}\approx(n+1)2^{n}$,
we have $P_{{\rm ue}}^{{\rm h}}(V_{0},p)\approx\frac{1}{n+1}$. This
statement can be made more precise. We also see that \[
P_{{\rm ue}}^{{\rm h}}(V_{0},0)=0=P_{{\rm ue}}(V_{0},0)\]
 and \[
P_{{\rm ue}}^{{\rm h}}(V_{0},1)=1-\frac{1}{\# V_{0}}=P_{{\rm ue}}(V_{0},1).\]

Now, let us consider the derivative of $P_{{\rm ue}}^{{\rm h}}(V_{0},p)$.
With simple algebra, we get \begin{eqnarray*}
\lefteqn{(n+1)^{2}\# V_{0}\frac{d}{dp}P_{{\rm ue}}^{{\rm h}}(V_{0},p)}\\
 & = & n(2-p)^{n-1}-\Bigl[n(2-p)^{n-1}-n(n-1)p(2-p)^{n-2}\Bigr]\\
 &  & +\Bigl[2n(1+p)^{n-1}+2n(n-1)p(1+p)^{n-2}\Bigr]-2n\\
 &  & -2n(n-1)p\\
 & = & 2n(1+np)\Bigl[(1+p)^{n-2}-1\Bigr]+n(n-1)p(2-p)^{n-2}\\
 &  & +2np.\end{eqnarray*}

In particular, we see that $\frac{d}{dp}P_{{\rm ue}}^{{\rm h}}(V_{0},p)>0$
for all $p\in(0,1)$; hence $P_{{\rm ue}}^{{\rm h}}(V_{0},p)$ is
increasing with $p$. Moreover, it is possible to show that $P_{{\rm ue}}^{{\rm h}}(V_{0},p)$
exhibits a nearly flat region on the interval $\Bigl[\frac{1}{\sqrt{n}},1-\frac{1}{\sqrt{n}}\Bigl]$.
This can be proved by considering that, for large $n$, the following
approximations hold (see Appendix II for demonstration):

\begin{eqnarray*}
P_{\mathrm{ue}}^{\mathrm{h}}\left(V_{0},\frac{1}{\sqrt{n}}\right)&\simeq&
   \frac{2^{n}}{\left(n+1\right)^{2}\# V_{0}}\left(1-\sqrt{n}e^{-\frac{\sqrt{n}}{2}-\frac{1}{8}}\right)\\
P_{\mathrm{ue}}^{\mathrm{h}}\left(V_{0},1-\frac{1}{\sqrt{n}}\right)&\simeq& 
  \frac{2^{n}}{\left(n+1\right)^{2}\# V_{0}}\left(1+2ne^{-\frac{\sqrt{n}}{2}-\frac{1}{8}}\right).
\end{eqnarray*}
 By using the approximation $\# V_{0}\simeq2^{n}/(n+1)$ we can obtain:

\begin{eqnarray*}
\lefteqn{P_{\mathrm{ue}}^{\mathrm{h}}\left(V_{0},1-\frac{1}{\sqrt{n}}\right)-P_{\mathrm{ue}}^{\mathrm{h}}\left(V_{0},\frac{1}{\sqrt{n}}\right)}\\
&\simeq&\frac{2^{n}}{\left(n+1\right)^{2}\# V_{0}}\left(2n+\sqrt{n}\right)e^{-\frac{\sqrt{n}}{2}-\frac{1}{8}}\simeq\frac{2n}{n+1}e^{-\frac{\sqrt{n}}{2}-\frac{1}{8}}.
\end{eqnarray*}
 Therefore 
\[P_{\mathrm{ue}}^{\mathrm{h}}\left(V_{0},1-\frac{1}{\sqrt{n}}\right)-P_{\mathrm{ue}}^{\mathrm{h}}\left(V_{0},\frac{1}{\sqrt{n}}\right)\rightarrow 0\]
for $n\rightarrow \infty$. 
Combined with the fact that
$P_{{\rm ue}}^{{\rm h}}(V_{0},p)$ is increasing, this confirms the existence
of a nearly flat region on the interval $\Bigl[\frac{1}{\sqrt{n}},1-\frac{1}{\sqrt{n}}\Bigl]$.
For example, if $n=509$, then $1/\sqrt{n}\simeq0.044$, and (\ref{eq12})
gives 
\[P_{{\rm ue}}^{{\rm h}}(V_{0},1/\sqrt{n})\simeq 0.001961, 
  P_{{\rm ue}}^{{\rm h}}(V_{0},1-1/\sqrt{n})\simeq 0.001972.\]

It is interesting to observe that the existence of a nearly flat region for
the probability of undetected error can be also proved, in general
terms, for any linear (or even non linear) code over the BSC. Demonstration
is given in Appendix III.

For Hamming codes, in particular, the existence of a nearly flat region in
the function $P_{\mathrm{ue}}^{\mathrm{BSC}}(H,p)$, given by 
(\ref{PueHBSC}), on the interval $\Bigl[\frac{1}{\sqrt{n}},1-\frac{1}{\sqrt{n}}\Bigl]$,
can be proved through similar arguments as those used above for the
VT codes. In this case, the derivative of the probability of undetected
error can be expressed as follows \cite[p. 44]{K}: 
\begin{eqnarray*}
\frac{dP_{{\rm ue}}^{{\rm BSC}}(H,p)}{dp}&=& n\left[(1-p)^{n-1}-(1-2p)^{(n-1)/2}\right] \\
&=& n\left\{ \left[(1-p)^{2}\right]^{(n-1)/2}-(1-2p)^{(n-1)/2}\right\}. 
\end{eqnarray*}
Since $\left(1-p\right)^{2}\geq\left(1-2p\right)$, it follows that
\[\frac{dP_{{\rm ue}}^{{\rm BSC}}(H,p)}{dp}>0\mbox{ for all }p\in(0,1),\]
and so $P_{{\rm ue}}^{{\rm BSC}}(H,p)$ is increasing with $p$. If we
consider the values of $P_{{\rm ue}}^{{\rm BSC}}(H,p)$ for $p=\frac{1}{\sqrt{n}}$
and $p=1-\frac{1}{\sqrt{n}}$, we can prove that, for large $n$,
the following approximations hold:

\begin{eqnarray*}
P_{\mathrm{ue}}^{\mathrm{BSC}}\left(H,\frac{1}{\sqrt{n}}\right) &\simeq & \frac{1+ne^{-\sqrt{n}-1}\left(1-\sqrt{e}\right)}{n+1}\\
P_{\mathrm{ue}}^{\mathrm{BSC}}\left(H,1-\frac{1}{\sqrt{n}}\right)&\simeq& \frac{1+ne^{-\sqrt{n}-1}}{n+1}.
\end{eqnarray*}
 Demonstration is given in Appendix II. It follows that

\begin{eqnarray*}
\lefteqn{P_{\mathrm{ue}}^{\mathrm{BSC}}\left(H,1-\frac{1}{\sqrt{n}}\right)-P_{\mathrm{ue}}^{\mathrm{BSC}}\left(H,\frac{1}{\sqrt{n}}\right)}\\
&\simeq &\frac{n}{n+1}e^{-\sqrt{n}-1}\left(1-1+\sqrt{e}\right) = \frac{n}{n+1}e^{-\sqrt{n}-\frac{1}{2}}.
\end{eqnarray*}
 Therefore 
\[P_{\mathrm{ue}}^{\mathrm{BSC}}\left(H,1-\frac{1}{\sqrt{n}}\right)-P_{\mathrm{ue}}^{\mathrm{BSC}}\left(H,\frac{1}{\sqrt{n}}\right)\rightarrow 0\]
for $n\rightarrow \infty$. Combined with the fact that
$P_{{\rm ue}}^{{\rm BSC}}(H,p)$ is increasing, this confirms the existence
of a nearly flat region on the interval $\Bigl[\frac{1}{\sqrt{n}},1-\frac{1}{\sqrt{n}}\Bigl]$
also in this case. In such region,
 \[
P_{{\rm ue}}^{{\rm BSC}}(H,p)\approx\frac{1}{n+1}.\]
 This is the same approximate value determined above for $P_{{\rm ue}}^{{\rm h}}(V_{0},p)$.
However, it is possible to verify that, for a fixed $n$, the extent
of the region where $P_{{\rm ue}}^{{\rm BSC}}(H,p)$ is almost constant
is larger than that where $P_{{\rm ue}}^{{\rm h}}(V_{0},p)$ is almost
constant.

Because of the lack of an explicit formula, it is not possible to
demonstrate analytically that the same nearly flat region appears also when
the Hamming code is applied over the Z-channel. However, the simulations described in the next section
indicate that this is the case.
So, assuming this, we can say that,
even keeping in mind the different meaning of $p$ over the symmetric
and the asymmetric channels, the curves of the probability of undetected
error for VT codes and Hamming codes of the same length over the Z-channel
and those for Hamming codes over the BSC are almost constant, and
practically superposed, in a wide region of the channel error probability.

\section{Performance simulation}

\label{sec:six}

In the previous section we have shown that the heuristic approach
provides a very good approximation for the case of small code lengths.
Testing reliability of the heuristic approximation for large lengths,
through a comparison with the exact results, is impossible, as the
exhaustive analysis becomes too complex just for $n>30$. For large
lengths, however, it can be useful to resort to a Monte Carlo like
method, that is, to develop a simulator. The simulator replicates
the behavior of a {}``real\char`\"{} system, and gives an estimate
of the unknown probability as the ratio between the number of undetected
errors and the number of simulated codewords.

A rule must be established to construct the code from the information
sequence. The simplest way to convert an information frame into a
codeword consists in applying a systematic encoding. Systematic VT
codes have been studied in \cite{AGF}. As reminded in Section \ref{sec:fourbis},
in a systematic code, every codeword consists of a $k$ bits information
vector and an $r$ bits parity check vector. In \cite{AGF}, a systematic
encoding procedure for VT codes of length $n$ and $r=\lceil\log_{2}(n+1)\rceil$
was given. This is, basically, the same that is obtained
with conventional Hamming codes (see Section \ref{sec:fourbis}).

The systematic encoding procedure described in \cite{AGF} is very
simple: the $k=n-\lceil\log_{2}(n+1)\rceil$ information bits are
set in the positions: \[
I=\left\{ 1,...,n\right\} \setminus\left\{ 2^{j}:j=0,1,...,\lceil\log_{2}(n+1)\rceil-1\right\} .\]
 $I$ defines a maximal standard information set for the VT code,
i.e., it ensures the value of $k$ is maximum. The remaining positions
are occupied by the parity check bits, whose values are determined
in such a way as to satisfy (\ref{eq:eq1}).

In general, the codewords of the systematic code, for a given value
of $n$, are a subset of those obtainable through the solution of
(\ref{eq:eq1}). On the other hand, it is evident that any codeword
of $V_{0}$ can be a codeword of the systematic code: in practice,
many information sequences can be encoded into more than one codeword
of $V_{0}$. As an example, for $n=10$, the information sequence
(011001) can be equivalently encoded into (1000110001) or into (0001110101).
When using the systematic code, one option should be chosen, when
necessary, in order to define the codewords uniquely. For our simulation
purposes, however, the goal is to generate the codewords of $V_{0}$
according to a uniform distribution. To this purpose, we do not adopt
any selection rule; on the contrary, when an information sequence
is randomly generated for transmission over the Z-channel, all its
possible encodings are considered. This way, simulation, that for
high values of $n$ necessarily corresponds to sampling a subset of
$V_{0}$, does not exhibit any {}``polarization effect\char`\"{}
and the simulated scenario strictly resembles that of the analytical
model (and the heuristic argument, in particular).

\begin{figure}[ht]
\centering \includegraphics[scale=0.75]{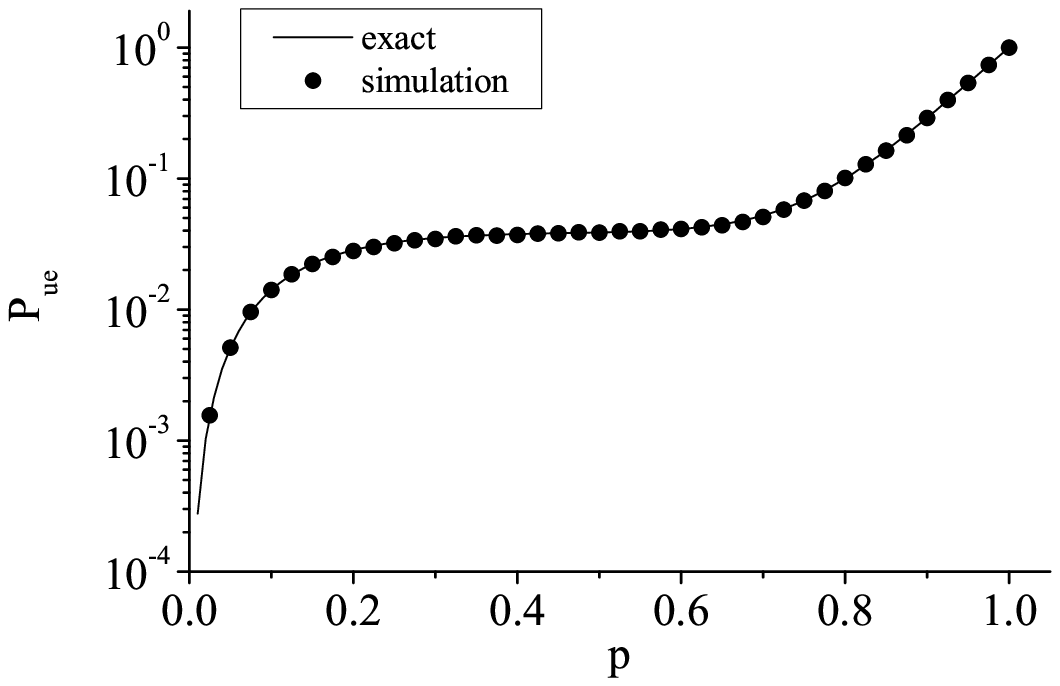}

\caption{Comparison between the simulated values and the exact curve of $P_{{\rm ue}}(V_{0},p)$
in the case of $n=25$.}

\label{fig:Sim25} 
\end{figure}

First, we have verified these conjectures by simulating the code with
$n=25$, that is the longest code for which we have presented before
the exact result; as shown in Fig. \ref{fig:Sim25}, the simulated
points are everywhere superposed to the exact curve.

Then, and most important, simulation has permitted us to study much
longer codes. We have analyzed lengths up to $n=509$ (that corresponds, according
with the systematic rule, to $k=500$). In order to ensure a satisfactory
statistical confidence level for the simulated $P_{{\rm ue}}(V_{0},p)$,
each simulation has been stopped after having found 50000 undetected errors.

\begin{figure}[ht]
\centering \includegraphics[width=8cm]{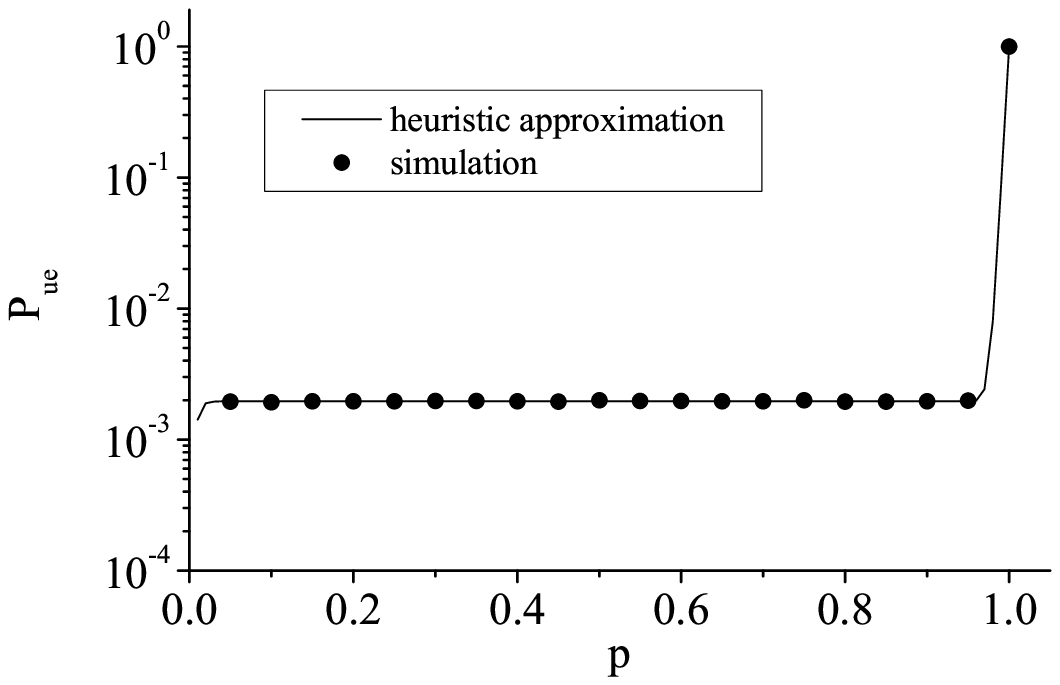}

\caption{Comparison between the simulated values of $P_{{\rm ue}}(V_{0},p)$
and the (heuristic) approximation $P_{{\rm ue}}^{{\rm h}}(V_{0},p)$
for $n=509$.}

\label{fig:Fig11} 
\end{figure}

The simulated curves for these long codes generally show a wide nearly
flat region, for intermediate values of $p$, as expected from the
heuristic analysis. Some examples of the numerical results obtained,
confirming the above considerations, are given in Table \ref{tab:LongCodes}.
For better evidence, in Fig. \ref{fig:Fig11} we have plotted the
heuristic approximation and the simulated values for $n=509$. We
see that the approximation is excellent also in this case. 

\begin{table}[ht]

\caption{{\small Examples of simulated values of $P_{{\rm ue}}(V_{0},p)$
for some values of $n$ and $p$} \label{tab:LongCodes}}

\begin{centering}
\begin{tabular}{cccccc}
\hline 
{\footnotesize $p$ }&
{\footnotesize $n=36$ }&
{\footnotesize $n=67$ }&
{\footnotesize $n=127$ }&
{\footnotesize $n=247$ }&
{\footnotesize $n=509$}\tabularnewline
\hline 
{\small 0.05}&
{\small 0.00643}&
{\small 0.00742}&
{\small 0.00647}&
{\small 0.00402}&
{\small 0.00195}\tabularnewline
{\small 0.1}&
{\small 0.01503}&
{\small 0.01261}&
{\small 0.00776}&
{\small 0.00403}&
{\small 0.00197}\tabularnewline
{\small 0.15}&
{\small 0.02088}&
{\small 0.01426}&
{\small 0.00785}&
{\small 0.00404}&
{\small 0.00196}\tabularnewline
{\small 0.2}&
{\small 0.02412}&
{\small 0.01463}&
{\small 0.00783}&
{\small 0.00403}&
{\small 0.00196}\tabularnewline
{\small 0.25}&
{\small 0.02573}&
{\small 0.01469}&
{\small 0.00780}&
{\small 0.00403}&
{\small 0.00197}\tabularnewline
{\small 0.3}&
{\small 0.02645}&
{\small 0.01465}&
{\small 0.00780}&
{\small 0.00404}&
{\small 0.00196}\tabularnewline
{\small 0.35}&
{\small 0.02677}&
{\small 0.01467}&
{\small 0.00782}&
{\small 0.00403}&
{\small 0.00196}\tabularnewline
{\small 0.4}&
{\small 0.02692}&
{\small 0.01479}&
{\small 0.00785}&
{\small 0.00404}&
{\small 0.00196}\tabularnewline
{\small 0.45}&
{\small 0.02692}&
{\small 0.01474}&
{\small 0.00780}&
{\small 0.00403}&
{\small 0.00197}\tabularnewline
{\small 0.5}&
{\small 0.02719}&
{\small 0.01469}&
{\small 0.00780}&
{\small 0.00402}&
{\small 0.00196}\tabularnewline
{\small 0.55}&
{\small 0.02729}&
{\small 0.01476}&
{\small 0.00781}&
{\small 0.00400}&
{\small 0.00195}\tabularnewline
{\small 0.6}&
{\small 0.02751}&
{\small 0.01467}&
{\small 0.00780}&
{\small 0.00402}&
{\small 0.00196}\tabularnewline
{\small 0.65}&
{\small 0.02785}&
{\small 0.01468}&
{\small 0.00781}&
{\small 0.00403}&
{\small 0.00196}\tabularnewline
{\small 0.7}&
{\small 0.02932}&
{\small 0.01468}&
{\small 0.00780}&
{\small 0.00401}&
{\small 0.00197}\tabularnewline
{\small 0.75}&
{\small 0.03402}&
{\small 0.01476}&
{\small 0.00779}&
{\small 0.00403}&
{\small 0.00196}\tabularnewline
{\small 0.8}&
{\small 0.04683}&
{\small 0.01553}&
{\small 0.00782}&
{\small 0.00404}&
{\small 0.00195}\tabularnewline
{\small 0.85}&
{\small 0.08159}&
{\small 0.01962}&
{\small 0.00793}&
{\small 0.00400}&
{\small 0.00197}\tabularnewline
{\small 0.9}&
{\small 0.17323}&
{\small 0.04481}&
{\small 0.00928}&
{\small 0.00405}&
{\small 0.00196}\tabularnewline
{\small 0.95}&
{\small 0.40865}&
{\small 0.19059}&
{\small 0.04674}&
{\small 0.00591}&
{\small 0.00195}\tabularnewline
\hline
\end{tabular}
\par\end{centering}
\end{table}

\begin{figure}[ht]
\centering \includegraphics[scale=0.75]{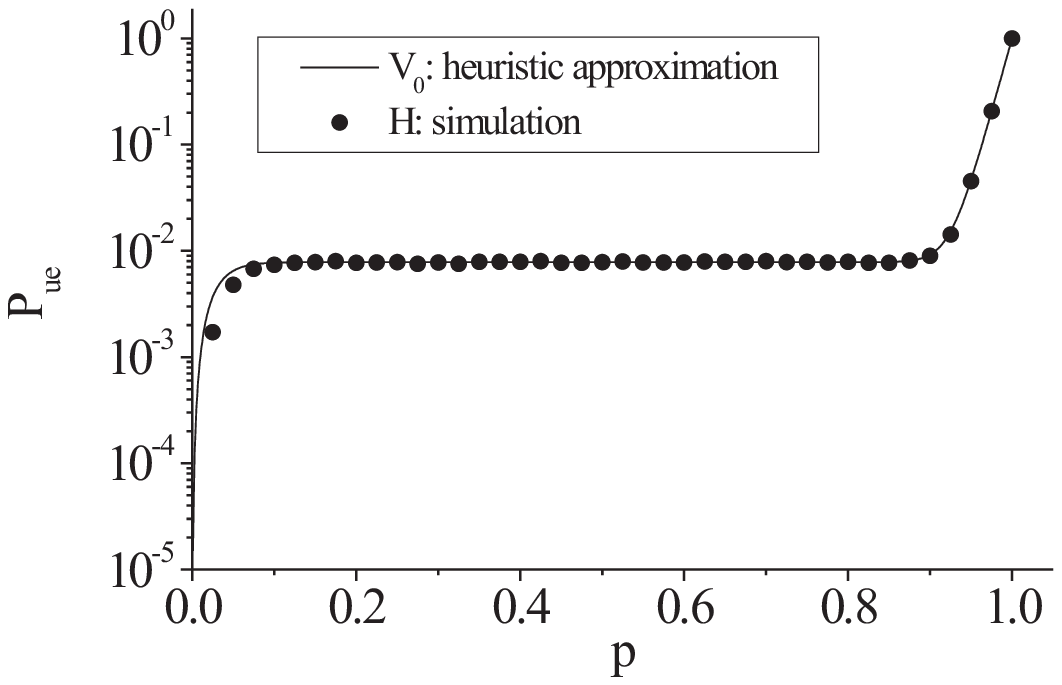}

\caption{Comparison between the simulated values of $P_{{\rm ue}}(H,p)$ and
the (heuristic) approximation $P_{{\rm ue}}^{{\rm h}}(V_{0},p)$ for
$n=127$.}

\label{fig:HvsV0127} 
\end{figure}

Finally, we can compare the performance of VT codes with that of Hamming
codes, with the same code length, most of all for demonstrating the
(quasi) coincidence of the nearly constant value. An example, for
$n=127$, is shown in Fig. \ref{fig:HvsV0127}: the continuous line
represents $P_{{\rm ue}}^{{\rm h}}(V_{0},p)$, while dots represent
some simulated points for $P_{{\rm ue}}(H,p)$. As expected, also
the latter curve exhibits a wide nearly flat region, and the value of both functions
are practically the same in this region. Moreover, this value is also
approximately equal to $1/(n+1)=0.0078125$ that, as proved in Section
\ref{sec:six}, provides the $P_{{\rm ue}}^{{\rm BSC}}(H,p)$ almost
everywhere, except for values of $p$ close to zero or close to one.

\section{Conclusion \label{sec:seven} }

This paper is a first attack on the problem of evaluating the undetected
error probability of Varshamov-Tenengol'ts codes. We have presented
some methods that allow us to obtain exact results (for short codes)
and heuristic and simulated approximate results (for long codes).
We have shown that the proposed heuristic approximation is excellent
for small $n$, and very good even for large $n$.

We have verified that the probability of undetected error is almost
constant in a wide region of values of the channel error probability,
and this region becomes larger and larger for increasing $n$. Such
a behavior is common to other codes, over the Z-channel, and can be
found even in the case of a generic, linear or non-linear, code over
the symmetric channel. Thus, we can conclude that, except for the
region of a channel error probability close to zero or one, the probability
of undetected error tends to assume the same value, approximately
equal to the reciprocal of the code length, independently of the code
and of the symmetry properties of the channel. Further work should be advisable to confirm these conclusions on other
codes. In regard to VT codes, though their error detection properties
seem disclosed from the analytical and numerical approaches proposed
in this paper, it remains a valuable task to find closed form expressions
for the quantities $A_{i,j}^{(0)}$, or even $A_{i,j}^{(g)}$ for
all $g$, in such a way as to be able to compute the undetected error
probability exactly for any code length.

\section*{Appendix I: On the heuristic approximation $P_{{\rm ue}}^{{\rm h}}(V_{0},p)$}

Let us consider (\ref{eq:eq3}), by assuming $g=0$ and replacing
the approximation (\ref{app2a}) for $j=i$ and (\ref{app2}) for
$j\neq i$; so, we get: 
\begin{eqnarray*}
\lefteqn{\# V_{0} P_{{\rm ue}}^{{\rm h}}(V_{0},p)}\\
 & = & \sum_{i=2}^{n}\frac{1}{n+1}\binom{n}{i}p^{i}\\
 && +\sum_{i=4}^{n}\sum_{j=2}^{i-2}\frac{1}{(n+1)^{2}}\binom{n}{i}\binom{i}{j}p^{j}(1-p)^{i-j}.
\end{eqnarray*}
 Through simple algebra, we have:
\begin{eqnarray*}
\lefteqn{(n+1)^{2}\# V_{0}P_{{\rm ue}}^{{\rm h}}(V_{0},p)} \\
&=& (n+1)^{2}\sum_{i=2}^{n}\frac{1}{n+1}\binom{n}{i}p^{i}\\
 && +(n+1)^{2}\sum_{i=4}^{n}\sum_{j=2}^{i-2}\frac{1}{(n+1)^{2}}\binom{n}{i}\binom{i}{j}p^{j}(1-p)^{i-j}\\
&=& (n+1)\sum_{i=2}^{n}\binom{n}{i}p^{i}+\sum_{i=4}^{n}\binom{n}{i}\sum_{j=2}^{i-2}\binom{i}{j}p^{j}(1-p)^{i-j}\\
&=& (n+1)\sum_{i=2}^{n}\binom{n}{i}p^{i} \\
 && +\sum_{i=4}^{n}\binom{n}{i}\Bigl[1-(1-p)^{i}-ip(1-p)^{i-1}\Bigr.\\
 && \qquad \Bigl.-ip^{i-1}(1-p)-p^{i}\Bigr] \\
&=& (n+1)\Bigl[(1+p)^{n}-1-np\Bigr]\\
 && + \Bigl[2^{n}-1-n-\binom{n}{2}-\binom{n}{3}\Bigr]\\
 && -\Bigl[(2-p)^{n}-1-n(1-p)\Bigr.\\
 && \Bigl.\qquad  -\binom{n}{2}(1-p)^{2}-\binom{n}{3}(1-p)^{3}\Bigr]\\
 && -np\Bigl[(2-p)^{n-1}-1-(n-1)(1-p)\Bigr.\\
 && \Bigl. \qquad -\binom{n-1}{2}(1-p)^{2}\Bigr]\\
 && -n(1-p)\Bigl[(1+p)^{n-1}-1-(n-1)p-\binom{n-1}{2}p^{2}\Bigr]\\
 && -\Bigl[(1+p)^{n}-1-np-\binom{n}{2}p^{2}-\binom{n}{3}p^{3}\Bigr]\\
&=& 2^{n}-(2-p)^{n}-np(2-p)^{n-1}\\
 && +2np(1+p)^{n-1}-2np-n(n-1)p^{2}.
\end{eqnarray*}

Hence 
\begin{eqnarray*}
\lefteqn{(n+1)^{2}\# V_{0}P_{{\rm ue}}^{{\rm h}}(V_{0},p)}\\
&=& 2^{n}-(2-p)^{n}-np(2-p)^{n-1}\\
&& +2np(1+p)^{n-1}-2np-n(n-1)p^{2}.
\end{eqnarray*}
This is the expression given in (\ref{eq12}) in Section \ref{sec:five}.

\section*{Appendix II: Approximate values of $P_{{\rm ue}}$ for $p=\frac{1}{\sqrt{n}}$
and $p=1-\frac{1}{\sqrt{n}}$.}

Let us consider, at first, the expression of $P_{\mathrm{ue}}^{\mathrm{h}}(V_{0},p)$,
given by (\ref{eq12}), for the approximate probability of undetected
error of VT codes over the Z-channel. For $p=\frac{1}{\sqrt{n}}$,
$P_{\mathrm{ue}}^{\mathrm{h}}(V_{0},p)$ takes the value

\begin{eqnarray}
\lefteqn{\left(n+1\right)^{2}\# V_{0} P_{\mathrm{ue}}^{\mathrm{h}}\left(V_{0},\frac{1}{\sqrt{n}}\right)} \nonumber\\
&=& 2^{n}-\left(2-\frac{1}{\sqrt{n}}\right)^{n}
-n\frac{1}{\sqrt{n}}\left(2-\frac{1}{\sqrt{n}}\right)^{n-1} \nonumber \\
 && +2n\frac{1}{\sqrt{n}}\left(1+\frac{1}{\sqrt{n}}\right)^{n-1} \nonumber\\
 && -2n\frac{1}{\sqrt{n}}-n\left(n-1\right)\frac{1}{n}.
\label{eq:Pueh_LeftPoint}
\end{eqnarray}

Considering that $\left(2-\frac{1}{\sqrt{n}}\right)^{n}=2^{n}\left(1-\frac{1}{2\sqrt{n}}\right)^{n}$,
we can adopt an approximate expression for such term. In fact, since
$0<\frac{1}{2\sqrt{n}}\leq\frac{1}{2}$, the Taylor expansion 
\[\mathrm{ln}\left(1-\frac{1}{2\sqrt{n}}\right)=-\frac{1}{2\sqrt{n}}-\frac{1}{8n}-\frac{1}{24n^{3/2}}-\cdots\]
can be used. This way, we obtain 
\begin{eqnarray*}
\left(2-\frac{1}{\sqrt{n}}\right)^{n} &=& 2^{n}e^{n\mathrm{ln}\left(1-\frac{1}{2\sqrt{n}}\right)}\\
 &=& 2^{n}e^{-\frac{\sqrt{n}}{2}-\frac{1}{8}+O\left(\frac{1}{\sqrt{n}}\right)} \\
 &\simeq& 2^{n}e^{-\frac{\sqrt{n}}{2}-\frac{1}{8}}.
\end{eqnarray*}
when $n\rightarrow\infty$.

Similarly, we can obtain $\left(2-\frac{1}{\sqrt{n}}\right)^{n-1}\simeq2^{n}e^{-\frac{\sqrt{n}}{2}-\frac{1}{8}}$
and $\left(1+\frac{1}{\sqrt{n}}\right)^{n-1}\simeq e^{\sqrt{n}-\frac{1}{2}}$;
so (\ref{eq:Pueh_LeftPoint}) can be rewritten as follows:

\begin{eqnarray*}
\lefteqn{(n+1)^{2}\# V_{0} P_{\mathrm{ue}}^{\mathrm{h}}\left(V_{0},\frac{1}{\sqrt{n}}\right)}\\
  &\simeq&  2^{n}-2^{n}e^{-\frac{\sqrt{n}}{2}-\frac{1}{8}}
    -\sqrt{n}2^{n}e^{-\frac{\sqrt{n}}{2}-\frac{1}{8}} \\
  && +2\sqrt{n}e^{\sqrt{n}-\frac{1}{2}}-2\sqrt{n}-n+1.
\end{eqnarray*}

Considering only the leading terms, we have 
\[P_{\mathrm{ue}}^{\mathrm{h}}\left(V_{0},\frac{1}{\sqrt{n}}\right)\simeq\frac{2^{n}}{\left(n+1\right)^{2}\# V_{0}}\left(1-\sqrt{n}e^{-\frac{\sqrt{n}}{2}-\frac{1}{8}}\right),\]
when $n\rightarrow\infty$.

We can adopt the same approach in order to obtain an estimate
of $P_{\mathrm{ue}}^{\mathrm{h}}(V_{0},1-\frac{1}{\sqrt{n}})$. We get

\begin{eqnarray*}
\lefteqn{(n+1)^{2}\# V_{0}P_{\mathrm{ue}}^{\mathrm{h}}\left(V_{0},1-\frac{1}{\sqrt{n}}\right)}\\
&=&  2^{n}-\left(1+\frac{1}{\sqrt{n}}\right)^{n}
    -n\left(1-\frac{1}{\sqrt{n}}\right)\left(1+\frac{1}{\sqrt{n}}\right)^{n-1}\\
&& +2n\left(1-\frac{1}{\sqrt{n}}\right)\left(2-\frac{1}{\sqrt{n}}\right)^{n-1}
   -2n\left(1-\frac{1}{\sqrt{n}}\right)\\
&& -n\left(n-1\right)\left(1-\frac{1}{\sqrt{n}}\right)^{2}.
\end{eqnarray*}

Using the approximations above, this can
be rewritten as follows:
\begin{eqnarray*}
\lefteqn{(n+1)^{2}\# V_{0} P_{\mathrm{ue}}^{\mathrm{h}}\left(V_{0},1-\frac{1}{\sqrt{n}}\right)}\\
 &\simeq& 2^{n}-\left(n+1-\sqrt{n}\right)e^{\sqrt{n}-\frac{1}{2}}\\
 && +2\left(n-\sqrt{n}\right)\left(2^{n}e^{-\frac{\sqrt{n}}{2}-\frac{1}{8}}-1\right)\\
 && -n(n-1)\left(1-\frac{1}{\sqrt{n}}\right)^{2}.
\end{eqnarray*}

Considering only the leading terms, we have 
\[P_{\mathrm{ue}}^{\mathrm{h}}\left(V_{0},1-\frac{1}{\sqrt{n}}\right)\simeq\frac{2^{n}}{\left(n+1\right)^{2}\# V_{0}}\left(1+2ne^{-\frac{\sqrt{n}}{2}-\frac{1}{8}}\right)\]
when $n\rightarrow\infty$.

A quite similar approach can be applied to the probability of undetected
error of Hamming codes, over the BSC, as given by (\ref{PueHBSC}).
In particular, we have: 
\begin{eqnarray*}
\lefteqn{P_{\mathrm{ue}}^{\mathrm{BSC}}\left(H,\frac{1}{\sqrt{n}}\right)} \\
&=& \frac{1}{n+1}+\frac{n}{n+1}\left(1-\frac{2}{\sqrt{n}}\right)^{\frac{n+1}{2}}-\left(1-\frac{1}{\sqrt{n}}\right)^{n}.
\end{eqnarray*}

As, for large $n$, $\left(1-\frac{2}{\sqrt{n}}\right)^{\frac{n+1}{2}}\simeq e^{-\sqrt{n}-1}$
and $\left(1-\frac{1}{\sqrt{n}}\right)^{n}\simeq e^{-\sqrt{n}-\frac{1}{2}}$,
this implies the following approximation:
\begin{eqnarray*}
P_{\mathrm{ue}}^{\mathrm{BSC}}\left(H,\frac{1}{\sqrt{n}}\right) &\simeq &
    \frac{1+ne^{-\sqrt{n}-1}-\left(n+1\right)e^{-\sqrt{n}-\frac{1}{2}}}{n+1}\\
  & \simeq &\frac{1+ne^{-\sqrt{n}-1}\left(1-\sqrt{e}\right)}{n+1}.
\end{eqnarray*}

At the point $p=1-\frac{1}{\sqrt{n}}$, instead we have: 
\[
P_{\mathrm{ue}}^{\mathrm{BSC}}\left(H,1-\frac{1}{\sqrt{n}}\right)=\frac{1}{n+1}\left[1+n\left(1-\frac{2}{\sqrt{n}}\right)^{\frac{n+1}{2}}\right]-n^{-\frac{n}{2}}
\]
 having taken into account that $(n+1)/2$ is always even. Moreover,
considering that $\left(1-\frac{2}{\sqrt{n}}\right)^{\frac{n+1}{2}}\simeq e^{-\sqrt{n}-1}$,
we have: 
\[
P_{\mathrm{ue}}^{\mathrm{BSC}}\left(H,1-\frac{1}{\sqrt{n}}\right)\simeq\frac{1+ne^{-\sqrt{n}-1}}{n+1}.\]

\section*{Appendix III: On the probability of undetected error of binary codes
over the symmetric channel}

Let $C$ be a binary $(n,M,d)$ code (it can be linear or non-linear).
By \cite[p. 44, Theorem 2.4]{K}, \[
P_{{\rm ue}}(C,p)=\frac{M}{2^{n}}\Bigl\{1+\sum_{i=d^{\perp}}^{n}A_{i}^{\perp}(1-2p)^{i}\Bigr\}-(1-p)^{n}\]
 where $A_{i}^{\perp}$ is the dual weight distribution (the MacWilliams transform of the weight distribution) of the code (for a linear code this is the weight distribution of the dual code) 
and $d^{\perp}$ the dual distance (that is, the least $i>0$ such
that $A_{i}^{\perp}\neq0$).

Therefore, 
\begin{equation}
\frac{d{P_{{\rm ue}}(C,p)}}{dp}=-\frac{M}{2^{n-1}}\sum_{i=d^{\perp}}^{n}iA_{i}^{\perp}(1-2p)^{i-1}+n(1-p)^{n-1}.
\label{der}\end{equation}

It is known that $A_{i}^{\perp}\ge0$ (see \cite[p. 16, Corollary 1.1]{K}).
Since $|1-2p|\le1$ for $p\in[0,1]$, we get \[
-(1-2p)^{i-1}\le|1-2p|^{i-1}\le|1-2p|^{d^{\perp}-1}\]
 for $i\ge d^{\perp}$. Further, it is known (see \cite[p. 14, Theorem 1.4]{K})
that \[
\frac{M}{2^{n-1}}\sum_{i=d^{\perp}}^{n}iA_{i}^{\perp}=n-A_{1}\le n.\]
 Hence, from (\ref{der}) we get 
\begin{eqnarray*}
\frac{d{P_{{\rm ue}}(C,p)}}{dp} & \le & n(1-p)^{n-1}+\frac{M}{2^{n-1}}\sum_{i=d^{\perp}}^{n}iA_{i}^{\perp}|1-2p|^{d^{\perp}-1}\nonumber \\
 & \le & n(1-p)^{n-1}+n|1-2p|^{d^{\perp}-1}.
\end{eqnarray*}

Similarly, we get 
\[
\frac{dP_{{\rm ue}}(C,p)}{dp}\ge n(1-p)^{n-1}-n|1-2p|^{d^{\perp}-1}.\]

The term $n(1-p)^{n-1}$ is close to zero for $p$ removed from zero
and one, for example for $1/\sqrt{n}\leq p\leq1-1/\sqrt{n}$.

The term $n|1-2p|^{d^{\perp}-1}$ is clearly small for $p$ close
to $1/2$. If $d^{\perp}$ is of some size, it is also small over
some \emph{range} around $p=1/2$. The bounds above show that $\frac{d{P_{{\rm ue}}(C,p)}}{dp}$
is also close to zero, and hence, \[
P_{{\rm ue}}(C,p)\approx P_{{\rm ue}}(C,1/2)=(M-1)/2^{n}\]
 over this range.

\end{document}